\documentclass[runningheads]{llncs}
\usepackage[T1]{fontenc}
\usepackage{graphicx}
\usepackage[none]{hyphenat}
\usepackage{hyperref}
\usepackage{color}
\usepackage{array}
\usepackage{multirow}
\usepackage{multicol}
\newcolumntype{C}{>{\centering\arraybackslash}m{0.25\textwidth}}
\newcolumntype{D}{>{\centering\arraybackslash}m{0.33\textwidth}}

\begin{document}
\title{Threat Me Right: A Human HARMS Threat Model for Technical Systems}
\titlerunning{Threat Me Right}

\author{Kieron Ivy Turk\inst{1}\orcidID{0000-0002-4705-4749} \and
Anna Talas\inst{1}\orcidID{0000-0001-7136-8839} \and
Alice Hutchings\inst{1}\orcidID{0000-0003-3037-2684}}
\authorrunning{K.I. Turk et al.}
\institute{Department of Computer Science and Technology, University of Cambridge \\
\email{\{kst36, at2008, ah793\}@cam.ac.uk}}
\maketitle              % typeset the header of the contribution
\begin{abstract}

Threat modelling is the process of identifying potential vulnerabilities in a system and prioritising them. Existing threat modelling tools focus primarily on technical systems and are not as well suited to interpersonal threats. In this paper, we discuss traditional threat modelling methods and their shortcomings, and propose a new threat modelling framework (HARMS) to identify non-technical and human factors harms. We also cover a case study of applying HARMS when it comes to IoT devices such as smart speakers with virtual assistants. 
\end{abstract}
\section{Introduction}
Secure systems require designers to actively identify threats to a system and create methods to prevent them. Threat modelling is the process of pre-emptively identifying and prioritising possible threats to a system or service. This process is structured using frameworks such as CIA and STRIDE to identify threats across a range of categories, often supplemented with a model such as DREAD to prioritise the interventions and mitigations.

Existing threat modelling tools are primarily designed around technical systems. For example, applying STRIDE  to a forum website may identify: users must be authenticated to prevent spoofing others; the site and database should be protected against tampering by hackers; users must be connected to posts through repudiation in case of harmful behaviours; personal information must be protected from disclosure; the site must avoid downtime from DDoS; and users must be distinct from moderators to prevent privilege escalation. Threat modelling can also be used for offline systems such as securing a bicycle: while a user may intuitively identify theft as a risk, CIA would invite them to consider the confidentiality of the owner's current location, the integrity of the bike frame and cables, or the availability of a bike with another person's lock attached.

While effective at evaluating the security threats in these traditional scenarios, threat modelling tools are not as well suited to interpersonal threats in scenarios such as domestic abuse. Prior research on technology-facilitated domestic abuse finds common threats are spying on and monitoring a partner~\cite{chatterjee18spywareipv,tseng20ipsinfidelityforums,ceccio23spydevices}, manipulation and gaslighting, harassment, and restricting access~\cite{woodlock17techdvandstalking,freed18AStalkersParadise,leitao21techabuseforums}. However, these do not fit the traditional threat modelling methodologies well. While surveillance threats can fit in Confidentiality from CIA or Information Disclosure in STRIDE, they are not the primary type of threat covered by these models. Furthermore, threats such as gaslighting do not fit well into any part of the prior frameworks. 

We argue a new threat modelling framework is required to properly encompass interpersonal threats and allow technologies to be designed against the abuser threat actor. We propose the HARMS threat modelling system, focusing on technology-facilitated abuse. This model has the potential to prevent a much wider category of misuse than explored by prior threat modelling systems, which would have aided with often-misused technologies such as AirTags~\cite{motherboard22police,nyt22airtags} and Ring Doorbells~\cite{stephenson23abusevectors,brown24safeguardingiot}. We demonstrate the use of this model by providing a case study of a smart speaker with a voice assistant and explore how this model can identify threats often missed by other threat modelling tools.

\section{Traditional Threat Modelling}

Threat modelling tools provide an organised methodology for identifying a range of threats to a system so that mitigations and interventions can be designed early on. These come in several forms: identifying threats through a series of prompts; prioritising threats; and providing a methodology for analysing a system.

\subsection{Identifying Threats}

\subsubsection{CIA}
The most universal threat modelling tool is the CIA triad--Confidentiality, Integrity, and Availability. CIA has formed the basis of most modern threat modelling, with references to the model dating back to the 1970s~\cite{neumann77cia}. Using the CIA triad ensures that system designers identify a broader range of threats early on, rather than focusing on the more obvious issues that arise from brainstorming.

\subsubsection{STRIDE}
An evolution of the CIA triad introduced by Microsoft~\cite{shostack18stride} is STRIDE-- Spoofing, Tampering, Repudiation, Information disclosure, Denial of service, and Escalation of privilege. This framework aims to cover a broader range of threats than the CIA triad while still including all of the core concepts; for example, confidentiality and information disclosure cover the same group of threats. STRIDE is widely used in industry for security systems.

\subsection{Prioritising Threats}

\subsubsection{DREAD}
While prior tools aim to assist in the threat discovery and exploration component of threat modelling, they do not provide a measure of the impact of different threats. To achieve this, Microsoft~\cite{howard02dread} introduced DREAD--Damage, Reproducibility, Exploitability, Affected users, and Discoverability. For each threat identified (e.g. using STRIDE), each DREAD component is scored from 1--10 and the sum is used to measure the overall risk posed by a threat vector.

\subsection{Alternative Modelling Processes}

\subsubsection{PASTA}
The Process for Attack Simulation and Threat Analysis (PASTA) consists of seven stages defined by Ucedav\'elez and Morana~\cite{ucedavelez15pasta}. These include defining objectives and scope, decomposing the application, analysing threats, vulnerabilities, and attacks, then concluding with risk and impact analysis. This end-to-end methodology is more complete and thorough than STRIDE and similar tools, making it a popular choice for non-security-focused companies.

\subsubsection{LINDDUN}
Instead of focusing on system threats, LINDDUN~\cite{wuyts20linddun}  provides a privacy-centric threat modelling system. The methodology consists of privacy threat types, threat trees, and three modelling methods. The threat types cover a wide range of privacy issues, including linking, identifying, non-repudiation, detecting, disclosure, unawareness, and non-compliance. This model is less technical, allowing a broader range of threat types to be identified.

\subsection{Shortcomings}
Existing threat modelling tools provide comprehensive coverage of technical threats to technical systems. However, these tools do not always apply in interpersonal abuse situations, where an abuser may use technology as intended, but in a way that inflicts harm on another. 
%However, most do not explore any form of human harms to technical systems, which
Interpersonal abuse poses a threat to users instead of the system. This causes issues when attempting to threat model systems around non-technical adversaries. For example, when considering technology-facilitated domestic abuse, Slupska and Tanczer~\cite{slupska2021threatmodellingipv} defined their own threat model as existing tools did not cover the observed threats. Their model includes ownership-based access, account and device compromise, harmful messages, exposure of information, and gaslighting. STRIDE only covers two of these points (compromising accounts comes under \textbf{E}scalation of privilege, and exposure of \textbf{I}nformation). 
To properly defend technical systems against human harms, we need to define a novel threat modelling system that covers threats to users rather than systems that can be used in conjunction with technical threat modelling tools.

\section{The Human HARMS Threat Model}
Existing threat modelling tools focus on technical threats while overlooking the human factors that can harm users in shared-access adversarial environments. We propose the HARMS threat modelling system to identify non-technical, human-factors harms to technical systems that fills this gap. The HARMS model contains five elements: \textbf{H}arassment, \textbf{A}ccess and infiltration, \textbf{R}estrictions, \textbf{M}anipulation and tampering, and \textbf{S}urveillance. These are summarised in Table~\ref{tab:harms-summary} and detailed throughout this section.

\begin{table}[t]
    \centering
    \caption{Summary of the Human HARMS Model}
    \begin{tabular}{|D|D|D|} \hline
        Term & Definition & Examples \\ \hline \hline
        Harassment & Causing distress through interactions & Sending hateful messages or playing loud sounds\\ \hline
        Access/Infiltration & Obtaining or extending access & Increasing own privileges, or adding an external user to a system\\ \hline
        Restrictions & Reducing access of existing user & Removing legitimate user's access, or inhibiting specific functionality\\ \hline
        Manipulation/Tampering & Controlling other users & Blackmailing users with information from the system, or creating fake evidence\\ \hline
        Surveillance & Observing others without their knowledge & Using cameras and microphones to observe users, or investigating logs of past activity\\ \hline
    \end{tabular}
    \label{tab:harms-summary}
\end{table}

\subsubsection{Harassment}
Malicious users may harass other users of the system. This may manifest as sending harmful messages, through standard messaging channels or subtle text inputs such as transaction messages. Users may also control devices to cause a disturbance by playing loud sounds, remotely toggling lights, or otherwise activating device functionalities. Key points for consideration include:

\begin{itemize}
    \item What channels are provided that could be used to send harmful messages?
    \item What features could be used to cause a disturbance? % speakers, flashing lights, etc
\end{itemize}

\subsubsection{Access and Infiltration}
Multi-user devices are commonplace, especially in smart homes where all residents want to be able to control devices. Researchers have found the majority of people in intimate relationships share access to their mobile phones~\cite{doerfler2024privacyvstransparency}. In these scenarios, malicious users can infiltrate systems they do not already have access to by learning existing user passwords, coercing access, or leveraging access to shared devices to authorise their own account on a system. These users may also attempt to use their access to allow third parties to infiltrate the system, sharing credentials or adding additional accounts to a system. Where users have limited access to a system, they may attempt to expand their control and gain more power over other users. Finally, some systems provide an inherent imbalance to certain users: for example, the user who sets up an IoT device often has more control over the system than a user with shared access. By being the user who configures the device, they have inherently expanded control over the system. The threat model may consider:

\begin{itemize}
    \item Can users gain unauthorised access to devices?
    \item Can users share access to unauthorised parties?
    \item Can users expand their control over a system or abuse access control to permanently have more access than other users?
\end{itemize}

\subsubsection{Restrictions} %(\& Escalation)}
Users in abusive home environments often attempt to gain a power imbalance by having more access than others in the system. This is often achieved by restricting access for other users, reducing their control over a device or system and impeding access to certain functionality. In some cases, abusive users may completely revoke access, despite having legitimate authorised access previously. Key points related to restrictions include:

\begin{itemize}
    \item Can one user restrict another user's capabilities?
    \item Can legitimate users have their access removed?
    \item Are there power imbalances through different user types? Are all users able to be assigned and maintain the correct privileges?
\end{itemize}

\subsubsection{Manipulation and Tampering}
Malicious users may abuse a system to manipulate others in various ways. They may \textit{gaslight} users by tampering with settings and data in the system and blame others, or perform actions that cannot be attributed to themself. They may use information gained through surveillance to blackmail or coercively control other users. Key considerations are:

\begin{itemize}
    \item Can users gaslight others through device features or information edits?
    \item Can information be obtained through the system for use in coercive control?
\end{itemize}

\subsubsection{Surveillance}
Users can abuse a range of system functionality to surveil others. This includes observing day-to-day activities, such as recording video or audio or monitoring location. Surveillance may also concern the use of a system, using logs, history, and other records to monitor another user's activity. Having access to a shared system may allow a user to obtain information about other users, such as personal information, schedules and events, and in some cases access to media including images and videos. Key points relating to surveillance include: 

\begin{itemize}
    \item Can the system be used to spy on users during day-to-day life?
    \item Can users see others' activity through the system? Are there any logs or history that could be used for this?
    \item Can personal data be accessed by shared users?
\end{itemize}

\subsection{Categorising Threats}
When considering the HARMS model, it is important to note that many identified threats do not fall into a single component. For example, doxxing someone by obtaining personal information and releasing it online constitutes both harassment and surveillance, although it is not a clear threat from either category individually. In particular, complex threats can constitute many components of the model. An adversary who observes a smart lock PIN code from outside the building, uses it to enter the house without authorisation, and then uses their building access to plant a hidden camera which can be used to obtain information for blackmail has combined harassment, infiltration, manipulation, and surveillance in one attack. Any threat analysis using HARMS must look at the potential to combine threats in this manner, rather than focusing on single category threats only.

Similar combinations of components can be observed with other models, such as STRIDE. A more technical adversary may install a keylogger (information disclosure), observe an email password and use it to later log in as the user (escalation of privilege), lock the user out of their account (tampering and denial of service), then perform actions on linked accounts (spoofing) whilst deleting evidence of their activities (repudiation). This would cover all six components in a single attack, whilst being a very feasible attack for a malicious insider who wants to cause harm to a company.

\section{Case Study: Smart Speaker}
Internet of Things (IoT) devices are often used for interpersonal harms~\cite{tanczer18bigbad}, and there are many IoT devices that would benefit from our model. We present an example of using the HARMS threat model to analyse a smart speaker with a voice assistant (such as an Amazon Echo or a Google Home): a device that is often connected to other smart home devices and can be controlled by talking to the built-in virtual assistant. We additionally analyse the system using STRIDE, and compare the results of the two methods.

\subsection{HARMS Analysis}
\subsubsection{Harassment}
Smart speakers provide functionality to read texts from a connected phone, or to announce messages sent from a system user. Both of these provide channels to have harmful messages read aloud to people in the vicinity. Additionally, the speaker can be used to play loud music or to set alarms late a night to cause a disturbance. The interconnectivity of these devices also allow them to control other devices in the household to cause a disturbance, such as flickering nearby lights or changing the thermostat to an uncomfortable temperature.

\subsubsection{Access and Infiltration}\label{ssec:speaker-harms-access}
Smart speakers are connected to a primary user, and any spoken commands will be performed through their account. This allows a user in the proximity of the speaker to infiltrate the owner's account to send messages, make purchases, or control their schedule. Spoken commands could also be used to access connected devices that the malicious user does not already have access to.

\subsubsection{Restriction}
If the adversarial user sets up and configures a device, they have full control over how other users get added to a system. In this case, they could set other adults in the system to have child accounts, so that they are able to impose parental controls and restrict their interactions with the speaker.

\subsubsection{Manipulation and Tampering}
Smart speakers allow interactions with connected calendars, which allows users to change or remove any scheduled events and make users miss meetings and appointments. They could also delete alarms or add new routines to gaslight the user into believing they are disorganised and making many errors. Furthermore, connected devices can be controlled remotely or automatically through routines, causing changes that the user does not understand and making them doubt their memory or even their sanity.

\subsubsection{Surveillance}
Many smart speakers provide ``drop-in'' features, allowing one user to call the device (without a requirement for users on the other end to pick up) and then listen in on the environment around the speaker. Users can either verbally query the device or look at logs to learn information about other users' routines, such as reminders and calendar events. Finally, the query history on the device allows users to see what others have asked the device, which is a likely source of personal information leaks.

\subsection{STRIDE Analysis}
\subsubsection{Spoofing}
Users that interact with a device are able to act as the device owner, providing an avenue for impersonation attacks. Certain devices are also vulnerable to voice commands using a carefully targeted laser~\cite{sugawara2020light}, allowing impersonation attacks without requiring access to the same physical space as the speaker.

\subsubsection{Tampering}
An experienced technical user could modify the hardware of the smart speaker to remove certain privacy features. For example, the hardware mute button could be disconnected to allow a user with access to listen constantly. Additional hardware could be covertly installed inside of the speaker body, such as externally controlled cameras and microphones.

\subsubsection{Repudiation}
Attackers can access the logs on devices and modify them to shift blame. Actions can be hidden from other users to gaslight them, or logs could be theoretically edited to show a different user performing a query or command.

\subsubsection{Information Disclosure}
Users are able to use built-in microphones to listen to ongoing conversations in the vicinity of the smart speaker. They may be able to use these microphones to constantly listen to the environment. Smart speakers record a history of commands and voice recordings, which can be accessed by an attacker. Devices linked to the speaker may provide additional information to the attacker about the smart home setup.

\subsubsection{Denial of Service}
Users could be locked out of a smart speaker if the voice recognition is trained to an attacker's voice, or if the owner's account is removed from the device. An attacker with access to the physical device can remove the microphone to prevent use, or damage the networking hardware to prevent internet access.

\subsubsection{Escalation of Privilege}
Privilege escalation attacks largely overlap with the Access and infiltration category of the HARMS model (\S\ref{ssec:speaker-harms-access}). A more sophisticated adversary may additionally attempt to increase access by installing malicious software onto the device through fake firmware updates or by directly interfacing with hardware. Alternatively, the attacker may target the owner's credentials, using common attacks such as credential stuffing or brute force/dictionary attacks to gain access to the account.

\subsection{Comparison of Methods}
\begin{table}[t]
    \centering
    \caption{Comparison of Threat Modelling Systems}
    \begin{tabular}{|C|C|C|C|}
        \hline
        \multirow{2}{*}{HARMS Terminology} & \multicolumn{3}{c|}{Overlap with} \\
        & CIA & STRIDE & LINDDUN \\\hline
        Harassment & --- & --- & --- \\
        Access/Infiltration & --- & Spoofing, Escalation of Privilege & --- \\
        Restriction & Availability & Denial of Service & --- \\
        Manipulation & Integrity & Tampering & --- \\
        Surveillance & Confidentiality & Information Disclosure & Linking, Identifying, Detecting, Disclosure \\\hline
        --- & --- & Repudiation & Non-repudiation \\
        --- & --- & --- & Unawareness, Non-compliance \\\hline
    \end{tabular}
    \label{tab:tm-comparison}
\end{table}
Both of these threat modelling analyses highlight a wide range of possible attacks on a smart speaker that should be defended against. Certain issues are better covered by one model, such as harassment through harmful messages in \textbf{H}ARMS or interfering with hardware in S\textbf{T}RIDE. Other attacks are caught by both models, such as using the voice assistant to edit the owner's calendar and reminders (HAR\textbf{M}S and S\textbf{TR}IDE). This showcases the benefits of using multiple threat modelling systems to identify a broader range of attacks against a system.

Certain parallels between many threat modelling systems can be observed, demonstrated in Table~\ref{tab:tm-comparison}. For example, obtaining personal information can be seen in surveillance in HARM\textbf{S}, information disclosure in STR\textbf{I}DE, and linking, identifying, detecting, and disclosure in \textbf{LI}N\textbf{DD}UN. Whilst these are fundamentally the same overarching threat, each model provides its own background to the issue, from more technical data breaches to privacy invasions and spying.

\section{Future Work and Extensions}
Our model identifies a range of different interpersonal harms that can be performed using technical systems. This model can be aided with further understanding of the threat actors who enact the harms, the environments in which they are perpetrated, and by providing actionable interventions within each category.

The concept of threat actors refers to the model of the person attacking or exploiting a system. Definitions range from simple concepts of a hacker or a malicious insider to comprehensive identification of goals and motivations, capabilities and limitations, and available resources. There are few existing threat actor definitions for the interpersonal harm actors in this model, and extensive literature review or lived experience are required to obtain a proper understanding of each threat actor. A future extension to this research can identify the unique threat actors and the risks they pose.

An additional component to understand the threats posed are the environments in which technical systems are exploited for harm. Traditionally, we are concerned with servers housed in data centers or personal machines in offices and at home, which constrain the types of attack possible for a malicious actor. When considering interpersonal harms, devices which are kept at home, used in a shared living space or dormitory, a workplace, or which move with the user all have different threat surfaces which pose risks. Follow-up work analysing threat actors should similarly consider the environments in which devices can be misused.

Traditional threat modelling analysis lends itself well to countermeasure design, as each possible threat has a property it violates that must be protected. For example, the six elements of STRIDE violate authenticity, integrity, non-repudiation, confidentiality, availability, and authorisation respectively. This is partially true for our model, as access would compromise authorisation and surveillance compromises privacy, but other HARMS risks are difficult to identify broad intervention strategies for. A framework which facilitates intervention design for threats identified through HARMS would be a beneficial extension to this model.

\section{Conclusions}
In this paper, we have presented a novel human HARMS threat modelling system and demonstrated how it can be used to identify interpersonal threats for a smart speaker in an adversarial home environment. We have also discussed existing threat modelling tools to demonstrate the gap left by these methods that our model fills. Our model provides additional security and safety against the rising threats of human harms, including domestic abusers, malicious online users and trolls, and stalkers.

\begin{credits}
\subsubsection{\ackname} We thank our colleagues at the Cambridge Cybercrime Center for their feedback. This work was supported by the UK Engineering and Physical Sciences Research Council (EPSRC) grants EP/W032473/1 and EP/T517847/1 and the European Research Council (ERC) under the European Union’s Horizon 2020 research and innovation programme (grant agreement No 949127).

% \subsubsection{\discintname}
% The authors have no competing interests to declare that are relevant to the content of this article.
\end{credits}
\bibliographystyle{splncs04}
\bibliography{threatmodelling}

\end{document}